\documentstyle[epsf]{mn2e}
\begin{document}

\title{Mid-IR period-magnitude relations for AGB stars}

\author[I.S. Glass, M. Schultheis, J.A.D.L. Blommaert, R. Sahai, M. Stute, 
S. Uttenthaler]{I.S. Glass$^1$,  M. Schultheis$^2$, J.A.D.L. 
Blommaert$^3$, R. Sahai$^4$, M. Stute$^5$ and \and S. Uttenthaler$^3$\\
$^1$South African Astronomical Observatory, PO Box 9, Observatory 7935, South
Africa\\
$^2$Observatoire de Besan\c{c}on, 41bis, avenue de l'Observatoire, F-25000 
Besan\c{c}on, France\\
$^3$Instituut voor Sterrenkunde, K.U. Leuven, Celestijnenlaan, 200D, B-3001 Leuven, Belgium\\
$^4$NASA\/JPL, 4800 Oak Grove Drive, Pasadena, CA 91109, USA\\
$^5$Department of Physics, University of Athens, Panepistimiopolis, 
15784 Zografos, Athens, Greece}

\date{Submitted 2008;
accepted}

\maketitle

\begin{abstract}

Asymptotic Giant Branch variables are found to obey period-luminosity
relations in the mid-IR similar to those seen at $K_S$ (2.14$\mu$m), even at
24$\mu$m where emission from circumstellar dust is expected to be dominant.
Their loci in the $M$, log$P$ diagrams are essentially the same for the LMC
and for NGC\,6522 in spite of different ages and metallicities. There is no
systematic trend of slope with wavelength. The offsets of the apparent
magnitude vs.\ log$P$ relations imply a difference between the two fields of
3.8 in distance modulus. The colours of the variables confirm that a
principal period with log $P$ $>$ 1.75 is a necessary condition for
detectable mass-loss. At the longest observed wavelength, 24$\mu$m, many
semi-regular variables have dust shells comparable in luminosity to those
around Miras. There is a clear bifurcation in LMC colour-magnitude diagrams
involving 24$\mu$m magnitudes.

\end{abstract}


\section{Introduction}

The existence of near-infrared period-luminosity relations in LMC Asymptotic
Giant Branch (AGB) stars is now well-known. They have also been found in the
NGC\,6522 Baade's Window of the inner Bulge by Glass \& Schultheis (2003)
and have since been extended to globular clusters and other members of the
Local Group. The LMC and the NGC\,6522 fields differ markedly in metallicity
but contain large numbers of stars at approximately uniform distances,
making them highly suitable places for studying population differences.

Among the AGB stars, many of the semi-regular variables have small
amplitudes so that, in contrast to the Miras, they can reveal useful
distance information from single-epoch observations if their periods are
otherwise known. Infrared studies have the additional advantage that the
effects of interstellar absorption are minimized; $A_{\lambda}$ in the
mid-IR can be as low as $\sim A_V \times 0.04$ mag.

Examination of the period-luminosity relations of AGB stars offers the
possibility of gaining insight into the physics of their outer atmospheres.
Many of the semi-regulars exhibit multiple periodicities. The principal
peaks in their periodograms lie in the range 10-200 days, usually
accompanied by other peaks within a factor of two in period and sometimes
with additional periods about 10 times as long as the principal ones. The
origin of the long periods is not yet understood. At least one case is known
in which a variation of this kind starts as a momentary decrease in
amplitude and develops in width over subsequent cycles (Blanco 26; see Glass
\& Schultheis, 2002).

One of the most important aspects of the long-period variables is their high
mass-loss rates. Though Mira variables are major contributors of matter to
the interstellar medium, the ISOGAL survey in the galactic plane (Omont et
al 2003), which made use of the ISO infrared satellite, showed that many
other late-type giants also shed large quantities of dust. The [7] - [15] vs
[15] colour-magnitude diagrams that resulted from ISOGAL show a continuous
distribution of mass-losing stars from the `blue' end of the colour range to
the `red' (Glass et al.\ 1999). Because these stars are comparatively
numerous they rival the Miras in their total dust output. Using an objective
prism survey of the NGC\,6522 field by Blanco (1986) it was possible to show
that the mass-losing stars are all giants of later spectral type than about
M5 (Glass \& Schultheis 2002). Moreover, thanks to variability data
generated as a by-product of the MACHO, OGLE and similar
gravitational-lensing experiments, it is also known that they are nearly all
semiregular variables, with the addition of a few Miras (Glass \& Alves
2000; Alard et al.\ 2001; Glass \& Schultheis 2002).

The Spitzer infrared satellite has now surveyed the Large Magellanic Cloud
(SAGE; Meixner et al.\ 2006) and the NGC\,6522 field{\footnote{As part of
the programme ID 2345: `A Spitzer Survey of Mass Losing Stars in the
Galactic Bulge', P.I.\ R.\ Sahai} at mid-infrared wavelengths with
sensitivities unobtainable from the ground. Each of these surveys used the
IRAC camera at 3.6, 4.5, 5.8 and 8 $\mu$m and the MIPS camera at 24 $\mu$m.
The SAGE data have been publicly released; the NGC\,6522 data were reduced
by M. Stute (Uttenthaler et al., in preparation). Because the NGC\,6522 field
has been so well studied in the past, it was an obvious candidate for a
Spitzer survey. It has importance as a fiducial field for application to
more heavily extincted areas of the inner Bulge where visible region data
are impossible to obtain.

Schultheis, Glass and Cioni (SGC2004) compared the $JHK_S$ properties of
AGB stars in the Milky Way galaxy and Magellanic Clouds.  In summary,
they formed complete samples by selecting all stars with
$M_{K_{S,0}}$ $<$ $\sim$ --5.0 in the relevant areas from the 2MASS Catalog.
This limit is about 1.5 mag below the tip of the RGB. The objects they found
were carefully cross-identified with the MACHO survey at $r$ and $i$. Each
MACHO light curve was Fourier analysed to yield its periods and associated
amplitudes. The sizes of the two fields of interest in this study, NGC\,6522
at 1032 arcmin$^2$ and the LMC at 271 arcmin$^2$, were chosen to yield about
1800 stars from each. The objects were also investigated using 7 $\mu$m and
15 $\mu$m photometry from ISO, where available. However, the Spitzer
instruments now offer high photometric precision at the faint end and
therefore the possibility of continuing the ISOGAL study to stars with lower
mass-loss rates. Spitzer also extends the long-wavelength end of the survey
from 15 to 24 $\mu$m, where radiation in excess of that expected from the
photospheres should be more clearly apparent.

\section{The data}

The NGC\,6522 and LMC fields discussed here using Spitzer data are
sub-fields of those investigated by SGC2004. In the current work, the
NGC\,6522 vs.\ MACHO and LMC vs.\ MACHO lists from SGC2004 were
cross-identified with the Spitzer photometry. The IRAC and MIPS fluxes were
converted into magnitudes by taking zero-points in the 3.6, 4.5, 5.8, 8 and
24 $\mu$m bands of 2.809 $\times 10^8$, 1.797 $\times 10^8$, 1.15 $\times
10^8$, 6.41 $\times 10^7$ and 7.14$\times 10^6$ $\mu$Jy respectively. The
reddening of the NGC\,6522 field was assumed to be $E_{B-V} = 0.5$.

The distance modulus (dm) for the LMC was taken to be 18.5. Following trials
of several values for the dm of the NGC\,6522 field, 14.7 was adopted (see
section \ref{pmd}).

In the LMC our survey covers about 0.00154 times the area of SAGE (Blum et
al 2006). Because we only discuss objects listed in the MACHO survey, a
number of very red objects detected by Spitzer do not appear in our lists.
They range in 24$\mu$m magnitude from 8.5 to 15.3 and in [3.6] -- [24]
colour from 4.1 to 9.0. Many of them may be AGB stars with circumstellar
shells thick enough to depress their $r$ and $i$ fluxes to levels below the
MACHO detection threshold. At least two objects more luminous than typical
AGB stars are present in our sample. One is a listed supergiant (WOH S 286).
The most apparently luminous of all, at $M_{24}$ = --15.25 is an H$\alpha$
object (LHA 120-N 132E) which has a MACHO counterpart with log$P$ = 2.661
(it is too bright to appear in Fig \ref{Pvsmags}). It should be mentioned,
in view of the comparisons we make, that Uttenthaler et al. (in preparation)
have reduced the SAGE data using their methods and find no systematic
differences with the results of Blum et al (2006).

The NGC\,6522 field does not contain any objects more luminous than the
Miras nor, unlike the LMC, does it possess any AGB stars with carbon
chemistry. Saturation (the criterion being that the aperture and PSF fluxes
disagree at $>$ 7\% level) sets in at 320, 330, 2300 1200 and 220 mJy for
the IRAC/MIPS bands, corresponding at dm = 14.7 to magnitudes of
--7.34, --7.86, --10.45, --10.36, --3.98. Thus measurements of the brighter
stars of the NGC\,6522 field in the [3.6] and [4.5] bands, though included
in the diagrams, are likely to be somewhat affected. 

We have included a number of AGB stars from the Solar Neighbourhood for
comparison. Because local AGB giants are too bright to be measured by
Spitzer, synthetic IRAC photometry of a number of them has been derived by
Marengo, Reiter and Fazio (2008) from ISO-SWS spectra. Periods for these,
where available, have been taken from the GCVS as they stand. Hipparcos (The
New Reduction) parallaxes (post-September 2008) were obtained from the
CDS. Only stars with parallaxes $\pi$ $>$ 5$\sigma_{\pi}$ have been
included. Most of the stars that satisfied this criterion are M-type; only
two C stars appear.

\begin{figure}
\epsfxsize=8.5cm
\epsffile[14 14 375 526]{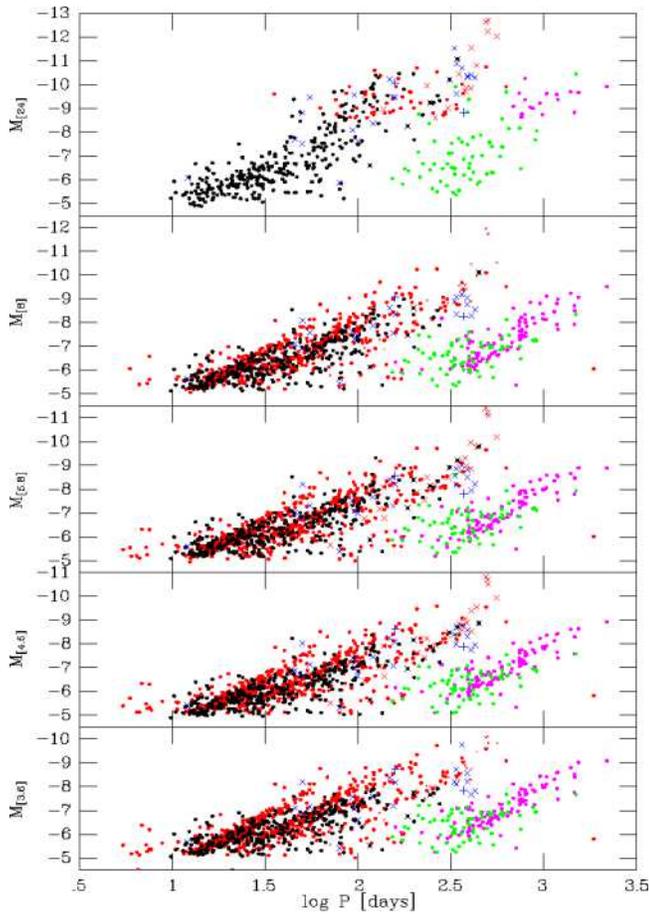}

\caption{log P vs.\ Spitzer magnitudes. Red dots are the LMC SRVs (principal
periods), red $\times$s the LMC Miras, magenta dots the LMC subsidiary long
periods and black and green the same respectively for NGC\,6522. The
principal $M$, log$P$ sequences are clearly visible. Note that the LMC data
are limited in sensitivity at 24$\mu$m. Bright NGC\,6522 points (longest
periods) suffer from saturation at 3.6 and 4.5$\mu$m (see text). Local AGB
stars from synthetic photometry by Marengo et al.\ (2008) are shown in blue
with $\times$ for M-types and + for C-types.}

\label{Pvsmags}
\end{figure}

\section{Period-magnitude diagrams}
\label{pmd}

Figure \ref{Pvsmags} contains the five $M$ vs.\ log$P$ diagrams for the two
fields, superimposed to show their essential similarity. Dots correspond to
the dominant periods. Miras are stars with MACHO amplitudes $>$ 1.0. Note
that the scatter in absolute magnitudes due to depth effects is much greater
in the NGC\,6522 than in the LMC field.

Versions of Fig \ref{Pvsmags} were prepared for several different values of
the dm of the NGC\,6522 field and compared visually. The fit of the
shortest-period SRVs to the corresponding sequence in the LMC appeared to be
the clearest discriminant and was best with dm = 14.7 to 14.8. Increasing or
decreasing the dm by 0.1 mag produced noticeable deviations. This value
appears to be similar to other determinations of the distance modulus of the
Galactic Centre based on AGB stars but is higher than those found using
other objects and methods. Groenewegen, Udalski \& Bono (2008) summarize the
current position and find dm = 14.50 based on RR Lyrae variables and
Population-II Cepheids. No clear explanation for this discrepancy has yet
emerged. It can be remarked that, because they are infrared-based, the
AGB-based determinations should be largely unaffected by interstellar
reddening.

The five Spitzer period-magnitude diagrams clearly show that the sequences
seen in the $K_S$, log$P$ diagrams by SGC2004 also appear at longer
wavelengths, though the LMC sample is limited at [24] by the sensitivity of
MIPS to $M_{[24]}$ brighter than --9. In spite of the presence of dust
shells around the longer-period stars at this wavelength, there is still
some tendency towards a relation. The fluxes from the shorter-period
variables remain photospheric. From $K_S$ to 8 $\mu$m essentially
the same $M$ -- log$P$ sequences are repeated. Though SGC2004 presented a
[7] vs log$P$ diagram and suggested that the sequences may persist at this
wavelength, they were not as obvious as here due to the smaller numbers of
stars, the limited depth of the photometry and possibly higher scatter in
the colours.

As expected from theoretical models, the LMC stars, being of lower
metallicity, tend to reach higher luminosities than the galactic ones. This
effect was previously remarked on by SGC2004 for the $K_S$ band. Groenewegen
and Blommaert (2005) point out also that the LMC contains higher mass stars
than the NGC6522 field. There are many additional subtle differences between
the $M$, log$P$ relations which are unfortunately difficult to quantify
because of the increased scatter due to depth effects in NGC\,6522.

In the case of the Solar Neighbourhood sample, 24$\mu$m magnitudes are not
available and absolute 25$\mu$m magnitudes derived from IRAS are plotted
instead.

\subsection{Slopes of the relations}

In order to determine whether the slopes of the period-magnitude relations
change with wavelength, straight lines were fitted to several of the
sequences visible in Fig.\ \ref{Pvsmags}. The periods were assumed free of
errors. The nomenclature of the sequences is based on that of Ita et al.
(2004). Only the LMC data were used for this part of the investigation since
they show the clearest separations. Not all of Ita's sequences are separate
enough to include in this analysis. Even among those that are included,
B$^+$ and B$^-$ are rather poorly defined and may be contaminated. The parts
of period-magnitude space used for fitting are shown in Fig.\
\ref{Pvsmags_spec} and the results are given in Table 1. The $K_S$ vs.\ $P$
data for LMC stars given by SGC2004 have been included in the analysis for
completeness.

As in Fig.\ \ref{Pvsmags}, stars with MACHO $r$ amplitudes greater than 1.0
were taken to be Miras. This level was chosen to include carbon Miras, which
tend to have lower amplitudes than M-types (see e.g. Glass \& Lloyd Evans
2003). Most Miras fall in the C region, though smaller amplitude stars are
also included. Objects with $J-K_S > 1.6$ were assumed to be carbon stars
(though this is not an infallible criterion) and were not included in the
linear fits.  Previous work such as that of Glass et al.\ (1987) and Feast
et al.\ (1989) suggested that O- and C-type Miras obey essentially the same
relations at $K_S$, though differing at $J$ and $H$ and consequently in
$M_{\rm Bol}$. The carbon Miras in Fig.\ \ref{Pvsmags_spec} seem to lie below
the M-star linear fit at $K_S$ but to be above it at longer wavelengths.
However, it should be noted that the sample is a small one and that the $K_S$
and IRAC data were not taken simultaneously.
   
Table 1 shows that the stars show little scatter ($\sigma$) around the
derived period-magnitude relations, especially in the A and B sequences. The
SRVs of these sequences have the smallest amplitudes; the C sequence,
however, contains some large-amplitude variables (Miras) and can be expected
to show more scatter. 

For each of the sequences A$^+$, A$^-$, B$^+$, B$^-$, C and D there is no
systematic change in slope with wavelength.

\begin{figure} 
\begin{center} 
\epsfxsize=8.5cm 
\epsffile[14 14 375 482]{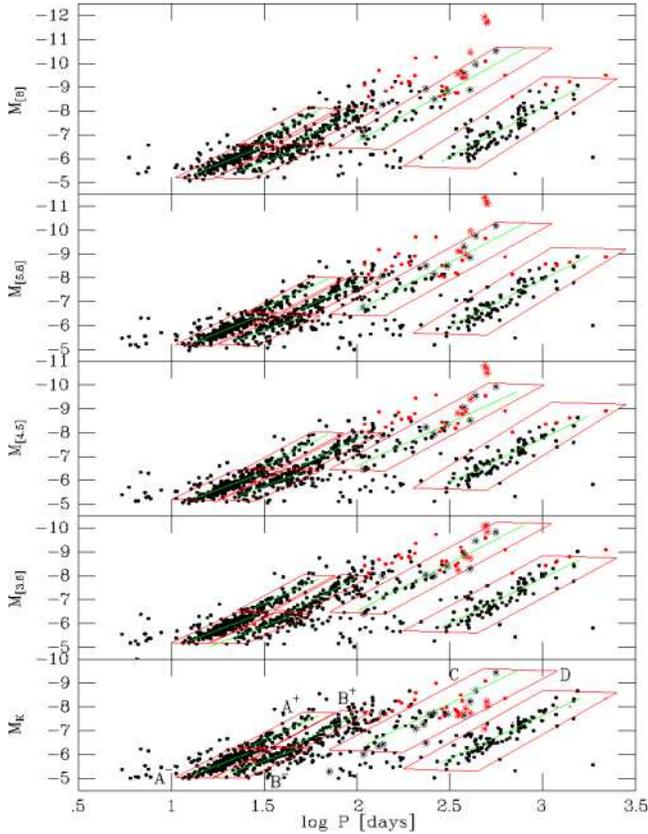} 

\caption{IRAC and 2MASS $K_S$ absolute mags vs.\ periods for the LMC sample.
Based on $J-K_S$ colour, the points representing oxygen-rich stars are black
and those for the carbon-rich are red. Mira variables are shown as
asterisks. The regions used for making the linear magnitude vs.\ log $P$
fits (to the O-rich stars only) are given in red. The coefficients of the
fitted lines in green are listed in Table 1.}

\label{Pvsmags_spec}
\end{center}
\end{figure}

\begin{table}
\caption{Linear fits to the $M$, log $P$ sequences}
\begin{tabular}{lllll}
Seq.   & Slope      & Const. term     & No & $\sigma$   \\
\multicolumn{5}{c}{$K_S$-band}\\
A$^-$ &-3.11$\pm$.11 & -1.67$\pm$.14 & 214 & 0.13   \\
A$^+$ &-3.46$\pm$.19 & -1.43$\pm$.29 &  67 & 0.15   \\
B$^-$ &-2.78$\pm$.14 & -1.57$\pm$.22 &  96 & 0.14   \\ 
B$^+$ &-3.85$\pm$.16 &  0.19$\pm$.29 &  70 & 0.13   \\ 
C     &-3.56$\pm$.29 &  0.86$\pm$.66 & 34  & 0.29   \\
D     &-3.61$\pm$.18 &  3.20$\pm$.51 & 90  & 0.29   \\
\multicolumn{5}{c}{3.6$\mu$m band}\\
A$^-$ & -2.99$\pm$.10 & -1.99  $\pm$.14 & 207 & 0.13 \\
A$^+$ & -3.50$\pm$.19 & -1.58  $\pm$.29 & 59 & 0.14 \\
B$^-$ & -2.89$\pm$.17 & -1.56  $\pm$.26 & 89 & 0.15 \\
B$^+$ & -4.05$\pm$.17 & 0.17   $\pm$.31 & 76 & 0.15 \\
C     & -4.00$\pm$.26 & 1.46   $\pm$.59 & 40 & 0.29 \\
D     & -3.78$\pm$.19 & 3.44   $\pm$.53 & 88 & 0.30 \\
\multicolumn{5}{c}{4.5$\mu$m band}\\                                                                        
A$^-$ & -2.95$\pm$.10 & -1.92  $\pm$.13 & 240 & 0.14 \\
A$^+$ & -3.75$\pm$.25 & -1.00  $\pm$.39 & 48 & 0.16 \\
B$^-$ & -2.90$\pm$.17 & -1.42  $\pm$.26 & 109 & 0.17 \\
B$^+$ & -3.81$\pm$.17 & -0.14  $\pm$.31 & 72 & 0.14 \\
C     & -3.62$\pm$.25 & 0.63   $\pm$.56 & 43 & 0.31 \\
D     & -3.74$\pm$.19 & 3.43   $\pm$.54 & 89 & 0.30 \\
\multicolumn{5}{c}{5.8$\mu$m band}\\
A$^-$ & -3.22$\pm$.11 & -1.68  $\pm$.14 & 230 & 0.14 \\
A$^+$ & -3.58$\pm$.20 & -1.36  $\pm$.32 & 53 & 0.14 \\
B$^-$ & -2.71$\pm$.18 & -1.74  $\pm$.28 & 101 & 0.18 \\
B$^+$ & -3.67$\pm$.20 & -0.46  $\pm$.36 & 83 & 0.19 \\
C     & -4.08$\pm$.25 & 1.542  $\pm$.57 & 41 & 0.30 \\
D     & -3.98$\pm$.21 & 4.01   $\pm$.58 & 87 & 0.33 \\
\multicolumn{5}{c}{8$\mu$m band}\\
A$^-$ & -3.11$\pm$.13 & -1.87  $\pm$.17 & 184 & 0.15 \\
A$^+$ & -4.21$\pm$.23 & -0.50  $\pm$.36 & 49 & 0.14 \\
B$^-$ & -2.80$\pm$.17 & -1.72  $\pm$.26 & 92 & 0.17 \\
B$^+$ & -3.75$\pm$.21 & -0.44  $\pm$.39 & 71 & 0.17 \\
C     & -4.35$\pm$.26 & 1.98   $\pm$.58 & 36 & 0.28 \\
D     & -4.15$\pm$.25 & 4.32   $\pm$.68 & 86 & 0.38 \\
\end{tabular}\\
Note: `No' is number of stars in sample. 
\end{table}

\section{Period-colour diagrams}
                                                                                                                                 
The colours of the samples are plotted against the principal and the very
long periods in Fig.\ \ref{Pvscols}. The most conspicuous trend is noticed
in the log $P$ vs.\ $K_{0,S}$ - [24] diagram, where the SRVs show a very
sharp onset of excess radiation at $P$ $\sim$ 60d. This is, in effect, an
exaggerated version of what is seen in the 15$\mu$m excess vs.\ log$P$
diagram of Alard et al.\ (2001). Noteworthy again is the tendency for many
SRVs to have very red infrared colours, comparable to, or even greater than,
those of the shorter-period Miras. It should be noted, however, from Fig
\ref{Pvsmags} that they are not necessarily as luminous.

The SRVs with secondary very long periods show clear evidence, particularly
at 24 $\mu$m, for the existence of dust shells. Unfortunately, the LMC
sample is limited to the most luminous cases by the sensitivity of MIPS.

\begin{figure}
\epsfxsize=8.5cm
\epsffile[14 14 375 555]{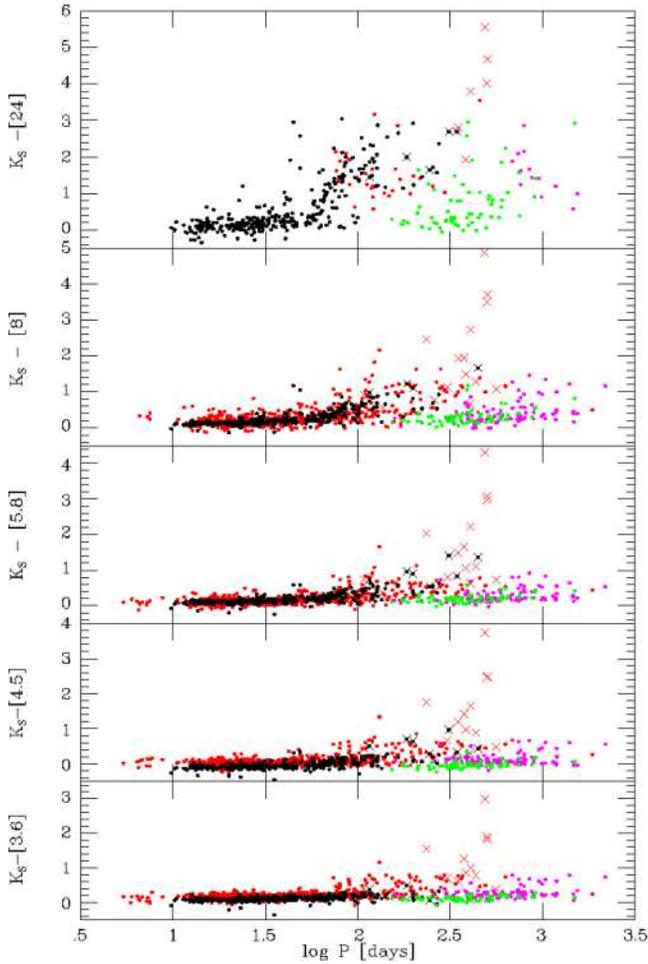}

\caption{log$P$ vs.\ $K_{S,0}$ - Spitzer colours. Red dots are the LMC SRVs
(principal periods), red $\times$s the LMC Miras, magenta dots the LMC
subsidiary long periods and black and green the same for NGC\,6522
respectively. The brightest stars in the NGC\,6522 field, such as the Miras,
suffer from saturation effects in the lower three diagrams.}

\label{Pvscols}
\end{figure}

\section{The [3.6] -- [24] vs.\ $M_{[24]}$ colour-magnitude diagram}

In this diagram (Fig \ref{marengo2}), AGB stars from both the LMC and
NGC\,6522 fields are shown, together with the local sample. As earlier,
LMC stars with $J-K_S > 1.6$ are taken to be C-type and the remainder
M-type.

The LMC points (red) are conspicuously bifurcated in the range --8.5 $>$ [24]
$>$ --11. Those on the right side of the fork (redder colours) are mainly
O-rich and comparable in position to the NGC\,6522 stars. The left fork
contains many carbon stars.

The carbon stars are clearly separated from oxygen-rich ones in the sense of
being less red in [3.6] -- [24] colour for a given $M_{[24]}$.  A similar
effect is seen whichever $m$ -- [24] colour is used on the abscissa but not
if say $M_{[8]}$ is used as the ordinate instead of $M_{[24]}$. Though the
difference may arise in part from the higher luminosities of the carbon
stars, there is a lack of emission from carbon-rich dust in the region
covered by the 24$\mu$m band. The IRAS [25] -- [60] vs.\ [12] -- [25]
colour-colour diagram (e.g. van der Veen and Habing, 1988) shows quite
clearly that galactic C-rich stars separate from O-rich ones at 25$\mu$m. In
the latter case, this part of the spectrum corresponds to the broad 18$\mu$m
silicate feature.

A similar trend was previously observed by SGC2004 in ISO data. In their
work, only 7 and 12$\mu$m photometry was available for the Magellanic Clouds
and 7 and 15$\mu$m for the NGC\,6522 field. The present use of identical
bands obviates any suggestion that the difference between the two fields
arises from the different spectral responses of ISO in the 12 and 15$\mu$m
bands.

Again, for local AGB stars, 25$\mu$m magnitudes from IRAS have been used.
These objects appear to lie slightly to the left of their NGC\,6522
counterparts. It is not clear whether this is due to the use of the IRAS
[25] band instead of the MIPS [24], to a calibration error or to a truly
physical effect such as higher luminosities in the Marengo et al.\ (2008)
sample. The two local C stars lie to the left of the M stars; larger samples
would of course be required to confirm this result.

A recent study of variables found by comparing two epochs of the entire SAGE
survey (Vijh et al.\ 2008) presents a [24] vs.\ [4.5] - [24] magnitude-colour
plot (their Fig. 4). This is, however, dominated by the class of `extreme
AGB stars'; SRVs would occupy mainly the bottom left corner (low-luminosity
and moderately red variables) and would be extremely numerous if all were
detected.

\begin{figure} 
\epsfxsize=8cm 
\epsffile[28 56 644 851]{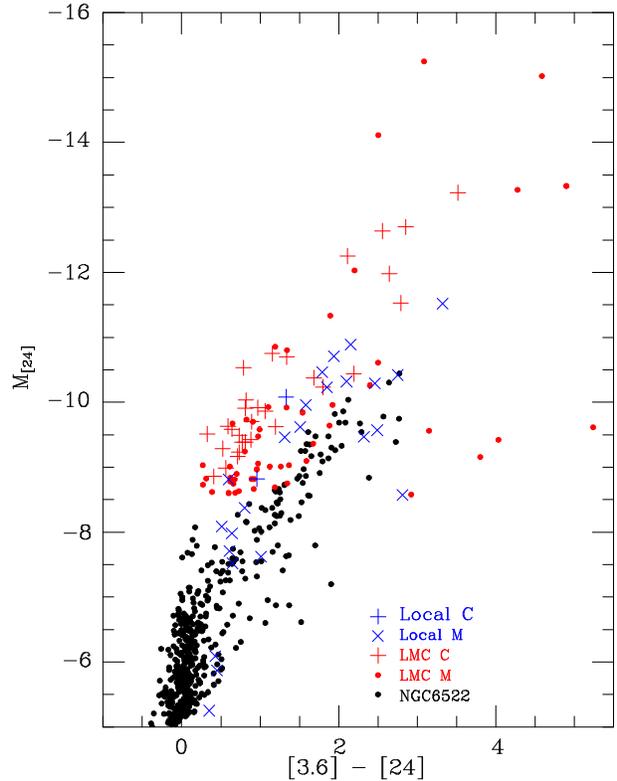}

\caption{[24] mag vs.\ [3.6] - [24] colour for LMC and NGC\,6522 AGB stars,
including non-variables. This diagram also incorporates nearby stars from
Marengo et al.\ (2008) with $\pi$ $>$ 5$\sigma_{\pi}$ using IRAS [25] mags
instead of Spitzer [24]. The LMC sample is divided into carbon stars (red +
symbols) and O-rich stars (red dots). Stars not detected by MACHO are
omitted. Note the bifurcation which we ascribe to a deficiency of
24--25$\mu$m flux in C-rich dust and possibly also to higher overall
luminosities. }

\label{marengo2}
\end{figure}

\section{Acknowledgments}

We thank Dr.\ M.\ Marengo for providing us with his detailed synthetic
photometry results. We acknowledge the use of the Simbad and Vizier on-line
resources of CDS, Strasbourg.  ISG's travel was supported by the CNRS/NRF
agreement. Dr.\ M.\ Groenewegen kindly commented on an earlier version of the
ms.

MSt is supported by the European Community's Marie Curie Actions - Human
Resource and Mobility within the JETSET (Jet Simulations, Experiments and
Theory) network under contract MRTN-CT-2004 005592. The data reduction of
the NGC6522 Spitzer data was carried out by MSt in collaboration with RS at
the Jet Propulsion Laboratory, California Institute of Technology, under a
contract with the National Aeronautics and Space Administration. RS received
partial support for this work through JPL/Caltech Spitzer Space Telescope GO
award MN0710076.

\end{document}